\documentclass[pra,aps,twocolumn]{revtex4}
\usepackage{epsf}
\usepackage{amsmath}
\usepackage{amssymb}

\def\<{\langle}
\def\>{\rangle}
\def \ket#1{{| #1 \rangle}}
\newcommand{\tr}{{\rm Tr}\,}

\begin{document}

\noindent {\bf Comment on ``Inseparability Criteria
for Continuous Bipartite Quantum States''}\\[12pt]
In a recent Letter [1] Shchukin and Vogel (SV) derived necessary
and sufficient conditions for a bipartite continuous variable (CV)
quantum state $\hat\rho$ to be positive under partial
transposition (PPT). The SV result shows a common basis of other
well-known CV inseparability criteria, some of which seemed
previously to be independent of partial transposition. The
argument used in Ref. [1] relates the condition of PPT of a
bipartite state to the positivity of a corresponding (infinite)
matrix of moments [see Eq. (12) of Ref. [1]]. From the latter, SV
derived an infinite series of inequalities [Eq. (20) of Ref. [1]],
and claimed that such series of inequalities provides a necessary
and sufficient condition for a state to be PPT (SV criterion). The
latter claim is  not correct, since inequalities (20) are
necessary but  not sufficient, as we prove by providing some
counterexamples. We propose an amended version of the criterion.

The SV conditions were formulated in terms of a hierarchy of
inequalities for the moments of creation and annihilation
operators $M_{ij}(\hat\rho)\equiv
M_{i_1i_2i_3i_4,j_1j_2j_3j_4}(\hat\rho)=\tr\big[(\hat{a}^{\dagger
i_{1}}\hat{a}^{i_{2}}\hat{b}^{\dagger
i_{3}}\hat{b}^{i_{4}})^\dagger (\hat{a}^{\dagger
j_{1}}\hat{a}^{j_{2}}\hat{b} ^{\dagger
j_{3}}\hat{b}^{j_{4}})\hat\rho\big]$. Here,  $i$ stands for the
multiindex $(i_{1},i_{2},i_{3},i_{4})$ (analogously for $j$), so
that we associate to $i$ an operator given by $\hat{a}^{\dagger
i_{1}}\hat{a}^{i_{2}}\hat{b} ^{\dagger i_{3}}\hat{b}^{i_{4}}$.
Moments form an infinite Hermitian matrix
$M(\hat\rho)=[M_{ij}(\hat\rho)]$. SV observed that the moments
calculated for the partially transposed state $\hat\rho^{\Gamma}$
(say, with respect to the second subsystem) correspond to moments
for $\hat\rho$ with reordered indices, i.e.,
$M_{i_1i_2i_3i_4,j_1j_2j_3j_4}(\hat\rho^{\Gamma})=
M_{i_1i_2j_3j_4,j_1j_2i_3i_4}(\hat\rho)$. in Ref. [1], Eq. (12),
it was found that $\hat\rho$ is PPT if and only if the (infinite)
matrix $M(\hat\rho^\Gamma)$ is positive semidefinite.

Let us recall Sylvester's criterion (see, e.g., [2]). For any
(possibly infinite) Hermitian matrix $\mathcal{M}$, let
$\mathcal{M}^{\bf r}$, ${\bf r}=(r_1,\ldots,r_N)$ denote the
submatrix which is obtained by deleting all rows and columns
except the ones labelled by $r_1,\ldots,r_N$. Moreover, let ${\cal
M}_N\equiv{\cal M}^{(1, 2, \ldots, N)}$, i.e., the submatrix
corresponding to the first $N$ rows and columns. Then Sylvester's
criterion can be formulated as follows: (i) ${\cal M}$ is positive
definite if and only if all its {\em leading principal} minors are
positive, i.e., $\det {\cal M}_N> 0$ for ${N}=1,2,\cdots$, while
(ii) ${\cal M}$ is positive semidefinite if and only if all its
{\em principal} minors are nonnegative, i.e., $\det {\cal M}^{{\bf
r}}\ge 0$ for any ${\bf r}\equiv(r_1,\ldots,r_N)$, $1\le r_1<
r_2<\ldots < r_{N}$, and ${N}=1,2,\ldots$.  In Ref. [1],
Sylvester's criterion (i) was used, leading the authors to
formulate incorrectly the following equivalent conditions:
\begin{align*}
\hat\rho~\textrm{ is PPT}~&\Leftrightarrow~\forall N:\quad \det
M_N(\hat\rho^\Gamma) \ge
0,\\
\hat\rho~\textrm{ is NPT}~&\Leftrightarrow~\exists N: \quad\det
M_N(\hat\rho^\Gamma)  <0,
\end{align*}
given by Eqs. (20) and (21), where NPT stands for  non-positive
under partial transposition. The SV entanglement criterion should
be based on Sylvester's criterion (ii) rather than (i), since we
deal with the request of positive semidefiniteness, which implies
that SV conditions should be modified to include all
subdeterminants of $M(\hat\rho^\Gamma)$ as follows:
\begin{align*}
\hat\rho~\textrm{ is PPT}~&\Leftrightarrow~\forall{\bf r}:\quad
\det M^{{\bf r}}
(\hat\rho^\Gamma) \ge 0,\\
\hat\rho~\textrm{ is NPT}~&\Leftrightarrow~\exists\,{\bf r}:
\quad\det M^{{\bf r}}(\hat\rho^\Gamma) <0.
\end{align*}
One can show that the original SV criterion does not reveal that
some exemplary states are NPT. Let us apply SV criterion to the
singlet state $\ket{\psi}= \frac1{\sqrt{2}}(\ket{01}-\ket{10})$,
e.g., in Fock basis. By using the same ordering of moments as in
Ref. [1], we find that the determinant $\det M_N(\hat\rho^\Gamma)$
is strictly greater than zero for $N=1,...,7$, and it vanishes for
$N\ge 8$. Thus, according to the criterion as stated originally in
Ref.~[1], one could draw the conclusion that the Bell state is
PPT. On the other hand, the inseparability of the Bell state is
revealed by  the modified criterion by choosing indices ${\bf r}$
corresponding to operators $1,\hat{a}\hat b$. The same problem
arises for other Bell states, for higher dimensional states, and
also for CV infinite-dimensional states. For example, one can
define a CV Bell state as a superposition of coherent states,
$\ket{\psi}\propto|\alpha, \beta\> - |-\alpha,-\beta\>$. We find
$\det M_N(\hat\rho^\Gamma)$ to be nonnegative for $N=1,...,7$ and
vanishing for $N\ge 8$. By contrast, the entanglement of the state
is revealed by selecting indices ${\bf r}$ corresponding to
operators $1,\hat b,\hat{a}\hat b$. In Ref.~[1], the authors
remark that it is possible to focus, for convenience (e.g. to
involve a lower number of moments), just on some principal - and
not necessarily leading - minors of $M(\hat\rho^\Gamma)$. Indeed,
the authors  detect the entanglement of the above mentioned CV
Bell state exactly with the choice of indices we listed. In the
same way they rederived the other already mentioned criteria of
entanglement in CV systems.

We stress that looking at principal minors - and not solely at the
leading principal minors - corresponds exactly to the spirit of
the amended criterion. It is clear that it is not only a matter of
convenience, but it is necessary to take into account the case of
singular matrices.

The authors thank R. Horodecki, P. Horodecki and M. Horo\-decki
for discussions. This work was supported by grants PBZ MIN
008/P03/2003 and 1~P03B 064 28 of KBN, and EU grants RESQ (IST
2001 37559), QUPRODIS (IST 2001 38877), EC IP SCALA.
\smallskip

\noindent Adam Miranowicz$^{1,2}$ and Marco Piani$^{1}$

\begin{small}

$^{1}$ Institute of Theoretical Physics and Astrophysics,\\
\hspace*{5mm} University of Gda\'nsk, 80-952 Gda\'nsk, Poland\\
\hspace*{3mm}$^{2}$ Institute of Physics, Adam Mickiewicz University,\\
\hspace*{5mm} 61-614 Pozna\'n, Poland.
\end{small}
\smallskip

Received 26 March 2006; published 2 August 2006

DOI: 10.1103/PhysRevLett.97.058901

PACS numbers: 03.67.Mn, 03.65.Ud, 42.50.Dv

\noindent [1]~E. Shchukin and W. Vogel, Phys. Rev. Lett. {\bf 95},
230502 (2005).

\noindent [2] {\em Matrix Theory and Applications}, edited by~R.
C. Johnson (American Mathematical Society, Providence, 1990).

\end{document}